\documentclass[draft]{agujournal2019}
\usepackage{url} %this package should fix any errors with URLs in refs.
\usepackage[inline]{trackchanges} %for better track changes. finalnew option will compile document with changes incorporated.
\usepackage{soul}
\usepackage{amsmath}
\allowdisplaybreaks 
\usepackage{threeparttable}
\usepackage{booktabs}
\usepackage{pdflscape}
\usepackage{colortbl}
\usepackage{multirow}

\usepackage{xr}
\draftfalse

\journalname{JGR: Machine Learning and Computation}

\begin{document}
%%%%%%%%%%%%%%%%%%%%%%%%%%%%%%%%%%%%%%%%%%%%%%%
% TITLE
%
% (A title should be specific, informative, and brief. Use
% abbreviations only if they are defined in the abstract. Titles that
% start with general keywords then specific terms are optimized in
% searches)
%
%%%%%%%%%%%%%%%%%%%%%%%%%%%%%%%%%%%%%%%%%%%%%%%

% Example: \title{This is a test title}

\title{Solar Energetic Particle Forecasting with Multi-Task Deep Learning: \texttt{SEPNET}}

%%%%%%%%%%%%%%%%%%%%%%%%%%%%%%%%%%%%%%%%%%%%%%%
%
% AUTHORS AND AFFILIATIONS
%
%%%%%%%%%%%%%%%%%%%%%%%%%%%%%%%%%%%%%%%%%%%%%%%

% Authors are individuals who have significantly contributed to the
% research and preparation of the article. Group authors are allowed, if
% each author in the group is separately identified in an appendix.)

% List authors by first name or initial followed by last name and
% separated by commas. Use \affil{} to number affiliations, and
% \thanks{} for author notes.
% Additional author notes should be indicated with \thanks{} (for
% example, for current addresses).

% Example: \authors{A. B. Author\affil{1}\thanks{Current address, Antartica}, B. C. Author\affil{2,3}, and D. E.
% Author\affil{3,4}\thanks{Also funded by Monsanto.}}

\authors{Yian Yu\affil{1}, Yang Chen\affil{1}, Lulu Zhao\affil{2}, Kathryn Whitman\affil{3}, Ward Manchester\affil{2}, Tamas Gombosi\affil{2}}

% \affiliation{1}{First Affiliation}
% \affiliation{2}{Second Affiliation}
% \affiliation{3}{Third Affiliation}
% \affiliation{4}{Fourth Affiliation}

\affiliation{1}{Department of Statistics, University of Michigan, Ann Arbor, MI, USA}
\affiliation{2}{Department of Climate and Space Sciences and Engineering, University of Michigan, Ann Arbor, MI, USA}
\affiliation{3}{NASA Space Radiation Analysis Group, Johnson Space Center, Houston, TX, USA}
%(repeat as many times as is necessary)

% Corresponding author mailing address and e-mail address:

% (include name and email addresses of the corresponding author. More
% than one corresponding author is allowed in this LaTeX file and for
% publication; but only one corresponding author is allowed in our
% editorial system.)

% Example: \correspondingauthor{First and Last Name}{email@address.edu}

\correspondingauthor{Yang Chen}{ychenang@umich.edu}

%%%%%%%%%%%%%%%%%%%%%%%%%%%%%%%%%%%%%%%%%%%%%%%
% KEY POINTS
%%%%%%%%%%%%%%%%%%%%%%%%%%%%%%%%%%%%%%%%%%%%%%%
% List up to three key points (at least one is required)
% Key Points summarize the main points and conclusions of the article
% Each must be 140 characters or fewer with no special characters or punctuation and must be complete sentences

% Example:
% \begin{keypoints}
% \item	List up to three key points (at least one is required)
% \item	Key Points summarize the main points and conclusions of the article
% \item	Each must be 140 characters or fewer with no special characters or punctuation and must be complete sentences
% \end{keypoints}

\begin{keypoints}
\item We propose a multi-task deep learning model, \texttt{SEPNET}, for SEP prediction. 
\item \texttt{SEPNET} predicts SEP events using summary statistics of solar flares, CMEs, and magnetic field measurements from SDO. 
\item \texttt{SEPNET}  offers earlier and more reliable alerts than traditional machine learning models, particularly with magnetic field measurements.
%\item \texttt{SEPNET} remains robust despite challenges with SEPs being rare events and false alarms in space weather prediction. 
\end{keypoints}

%%%%%%%%%%%%%%%%%%%%%%%%%%%%%%%%%%%%%%%%%%%%%%%
%
% ABSTRACT and PLAIN LANGUAGE SUMMARY
%
% A good Abstract will begin with a short description of the problem
% being addressed, briefly describe the new data or analyses, then
% briefly states the main conclusion(s) and how they are supported and
% uncertainties.

% The Plain Language Summary should be written for a broad audience,
% including journalists and the science-interested public, that will not have a background in your field.
%
% A Plain Language Summary is required in GRL, JGR: Planets, JGR: Biogeosciences,
% JGR: Oceans, G-Cubed, Reviews of Geophysics, and JAMES.
% see http://sharingscience.agu.org/creating-plain-language-summary/)
%
%%%%%%%%%%%%%%%%%%%%%%%%%%%%%%%%%%%%%%%%%%%%%%%

%% \begin{abstract} starts the second page

\begin{abstract}
 Solar energetic particle (SEP) events pose severe threats to spacecraft, astronaut safety, and aviation operations. Accurate SEP forecasting remains a critical challenge in space weather research as a result of their complex origins and highly variable propagation. In this work, we built \texttt{SEPNET}, an innovative multi-task neural network that jointly predicts future solar eruptive events, including solar flares and coronal mass ejections (CMEs) and SEPs, incorporating long short-term memory and transformer architectures that capture contextual dependencies. 
 \texttt{SEPNET} is a machine learning framework for SEP prediction that utilizes an extensive set of predictors, including the properties of solar flares, CMEs, and space-weather HMI active region patches (SHARP) magnetic field parameters. 
 \texttt{SEPNET} is rigorously evaluated on the SEPVAL SEP dataset \cite{whitman2025SEPVAL}, which is used to evaluate the performance of current SEP prediction models. The performance of \texttt{SEPNET} is compared with classical machine learning methods and current state-of-the-art pre-eruptive SEP prediction models. The results show that \texttt{SEPNET}, particularly with SHARP parameters, achieves higher detection rates and skill scores while maintaining the suitability for real-time space weather alert operations. Although class imbalance in the data leads to relatively high false alarm rates, \texttt{SEPNET} consistently outperforms reference methods and provides timely SEP forecasts, highlighting the capability of deep multi-task learning for next-generation space weather prediction.
\end{abstract}

\section*{Plain Language Summary}
Explosions on the Sun can send high-energy solar energetic particles (SEPs) into space. SEP events are a type of solar radiation storm that can affect astronauts, satellites, and high-latitude aircraft because the particles can damage the electronics and pose a safety risk to humans. In this work, we presented \texttt{SEPNET}, a new machine learning tool that uses solar eruptive events, including solar flares, CMEs, and solar magnetic field measurements, to predict when SEP events will occur. \texttt{SEPNET} uses artificial intelligence to learn from historical space weather data and provides more accurate early warnings than existing methods. \texttt{SEPNET} shows promise in helping scientists and decision-makers protect us from the risks of space weather. 

%Enter your Plain Language Summary here or delete this section.
%Here are instructions on writing a Plain Language Summary: 
%\url{https://www.agu.org/Share-and-Advocate/Share/Community/Plain-language-summary}
%%%%%%%%%%%%%%%%%%%%%%%%%%%%%%%%%%%%%%%%%%%%%%%
%
% BODY TEXT
%
%%%%%%%%%%%%%%%%%%%%%%%%%%%%%%%%%%%%%%%%%%%%%%%

%%% Suggested section heads:
% \section{Introduction}
%
% The main text should start with an introduction. Except for short
% manuscripts (such as comments and replies), the text should be divided
% into sections, each with its own heading.

% Headings should be sentence fragments and do not begin with a
% lowercase letter or number. Examples of good headings are:

% \section{Materials and Methods}
% Here is text on Materials and Methods.
%
% \subsection{A descriptive heading about methods}
% More about Methods.
%
% \section{Data} (Or section title might be a descriptive heading about data)
%
% \section{Results} (Or section title might be a descriptive heading about the
% results)
%
% \section{Conclusions}

\section{Introduction}
Solar energetic particle (SEP) events are transient releases of high-energy protons, electrons, and heavy ions accelerated during solar flares and coronal mass ejections (CMEs) \cite{hilberg1969radiation,iucci2005space}. These charged particles constitute significant radiation hazards to spacecraft electronics, astronaut safety, and high-latitude aviation operations \cite{Eastwood2017Economic,Whitman20235161}. As human activities extend beyond low Earth orbit, accurate real-time forecasting of SEP events has become increasingly vital, yet remains challenging due to their intermittent occurrence and the complex mechanisms underlying particle acceleration and interplanetary transport \cite{REAMES2004381,KIM2011747,desai2016large,klein2017acceleration}. In recent decades, SEP prediction has advanced through empirical, physics-based, and machine-learning methods, with the aim of balancing predictive accuracy with operational timeliness \cite{smart1979pps76,Opgenoorth2019Assess,Kasapis2022SW,ali2025forecasting}.

Traditional SEP models typically integrate physical understandings of particle acceleration at solar flares and CME-driven shocks with solar eruption observations through empirical relations or physics-based acceleration and transport simulations. Empirical models rely on statistical correlations derived from historical data to rapidly forecast SEP occurrence or intensity using flare, CME, and radio burst parameters, but may lack detailed physical interpretation \cite{smart1979pps76,Balch2008Update,Laurenza2009Atech}. Physics-based models simulate the fundamental processes of SEP acceleration and transport in the corona and heliosphere by coupling solar wind, CME shock evolution, and particle kinetics, often solving the transport equations and modeling diffusive shock acceleration \cite{Sokolov2004AJ,Luhmann2007295,Hu2017Modeling,Young2021Energetic,zhao2023cCLEAR, zhao2024solar}. Despite their interpretability and scientific value, these physics-based models tend to be computationally intensive, and there are still uncertainties in key input parameters such as seed particle populations and accurate CME/shock characteristics \cite{Tylka2006,NeergaardParker2012,Desai2020Properties}. For empirically driven forecasting, additional observational constraints, such as delays in coronagraph data acquisition, limited real-time radio observations, and imperfect knowledge of magnetic connectivity to the observing spacecraft, also pose challenges for operational deployment \cite{erickson1997bruny,Gopalswamy2005JA011158,richardson201425}. The trade-offs between physical completeness and operational practicality lead to a proliferation of varied model designs, each with advantages and limitations regarding forecast accuracy, timeliness, and interpretability \cite{Whitman20235161}.

The growing availability of diverse, multichannel, and multiwavelength solar observational data, together with advances in machine learning (ML) techniques, has spurred numerous ML-based approaches for SEP forecasting \cite{Kasapis2022SW,Whitman20235161, Dayeh2024SW}. ML models typically incorporate features such as solar flare characteristics, CME parameters, and photospheric magnetic field descriptors, such as the space-weather HMI active region patches (SHARP). These models have demonstrated competitive or superior predictive performance compared with traditional empirical or physics-based models, improving operational timeliness and accuracy. For instance, convolutional neural networks, support vector machines, and ensemble tree-based methods have been used to predict SEP occurrence probabilities and intensities by leveraging feature sets including flare X-ray flux, CME speed and width, and magnetic field proxies \cite{boubrahimi2017prediction,lavasa2021assessing,Kasapis2022SW}. Recent work by \citeA{Ji2025FeatureMapping} advances this field by proposing a novel framework that combines global feature mapping and multivariate time-series classification to enhance model interpretability and accuracy. 

This work is among the first to employ transformer architectures for SEP prediction in a multi-task deep learning setting. The use of transformers in heliophysics has previously been limited to solar flare forecasting \cite{Kaneda2022FlareTS,Abduallah2023FlareTransformer,Pelkum2024SolarTrans}, general-purpose representation learning given by the SuryaFM foundation model \cite{Roy2025Surya}, and continuum intensity prediction for solar active region emergence \cite{Tirona2026Forecasting}. Unlike conventional single-task learning frameworks, multi-task learning models jointly learn related prediction tasks by sharing latent representations, which improves generalization and mitigates overfitting, especially in data-constrained environments \cite{caruana1997multitask, Yu2017MTL,Crawshaw2020MultiTaskLW}. The inherently interconnected nature of solar eruptive phenomena, where flares, CMEs, and SEPs are physically and temporally coupled, naturally motivates multi-task learning approaches. To date, SEP prediction efforts have often treated SEP occurrence, flare forecasting, and CME characteristics as separate or sequential problems. However, joint modeling through multi-task learning can exploit shared underlying physics and temporal correlations, yielding more accurate and stable predictions.

A notable limitation in previous ML models was the limited size and diversity of available SEP event datasets, as well as the lack of a consistent benchmark set of validation periods to allow cross-model comparisons of performance. For example, \citeA{Kasapis2022SW} reported a predictive accuracy of approximately 0.72 using a modest dataset of 65 SEP events, highlighting the critical need for larger, curated datasets spanning recent solar cycles to enable more robust model training and validation. \citeA{Kasapis2022SW} also noted that a fair comparison between different model types was not feasible due to a lack of consistent underlying testing and training data.  Moreover, \citeA{Kasapis2024SEPMARP} extended this research by employing interpretable machine learning for SEP forecasting across solar cycles 23 and 24 using the merged SHARP-SMARP magnetic-parameter dataset, highlighting the value of a larger, multi-cycle magnetic-field dataset for SEP prediction.

The SEPVAL initiative established a collaborative, multi-year benchmark for SEP model validation by compiling and curating a dataset comprised of 33 SEP events and 30 non-event periods, involving model developers, operational stakeholders, and the space weather research community \cite{SEPVAL2023,whitman2024multi}. A detailed introduction to the SEPVAL dataset is provided in Section \ref{sec::DataPreprocessing}. Building upon the infrastructure developed to support SEPVAL, the CLEAR SEP Benchmark Dataset was created to provide an expanded dataset for scientific analysis and model training.  In this paper, we use the Operational version of the SEP data product from the CLEAR Center (\url{https://science.nasa.gov/clear/}), compiled through September 2025, including detailed records of solar flares, CMEs, and SHARP magnetic field parameters, along with a rigorously curated catalog of SEP events. In this dataset, the particle and detector background were identified and set to zero, leaving non-zero fluxes only during enhanced periods. SEP events were identified above background (indicated by an arbitrarily low threshold of 1e-6 pfu) and above multiple operational thresholds (e.g. $>$ 10 MeV $>$ 10 pfu). The version of the \texttt{FetchSEP} package used to generate the CLEAR SEP dataset is available at \citeA{fetchsep}. 

The extensive dataset underpins the training of a novel multi-task learning model, \texttt{SEPNET} designed to simultaneously predict SEP event occurrence and continuous flare and CME parameters. \texttt{SEPNET} employs shared neural network layers coupled with task-specific output heads, effectively capturing the latent interdependencies inherent in the three solar eruptive phenomena, flares, CMEs and SEPs. By treating flare and CME forecasting as auxiliary objectives, the model leverages these related physical signatures to regularize and enhance the primary task of SEP prediction. \texttt{SEPNET} integrates temporal dynamics through recurrent long short-term memory (LSTM) \cite{Hochreiter1997LSTM} and attention-based transformer \cite{vaswani2023attentionneed} architectures, enabling the exploitation of sequential dependencies in solar observations and, thereby, enhancing predictive capabilities compared to traditional single-task classifiers.

The remainder of this paper is organized as follows. Section \ref{sec:Method} details the methodology, including data preparation, preprocessing steps, and the development of the \texttt{SEPNET} model. In this section, we describe the feature selection procedure and the strategies employed for model training and evaluation. In Section \ref{sec:results}, we present the results of applying the \texttt{SEPNET} model and the upgraded version (\texttt{SEPNET-TS}) that incorporates temporal information to the SEPVAL data set, including a comparative analysis with conventional machine learning classifiers and real-time operational prediction of SEP (\texttt{SEPNET-O}). Finally, Sections \ref{sec:discussion} and \ref{sec:conclusion} discuss and summarize the main findings and conclusions of this study, and directions for future research. Additional details and supplementary material are provided in the Supporting Information. 

\section{Methodology}\label{sec:Method}
This section describes the data preparation, preprocessing, and model development used in this study. Section \ref{sec::Data_Prep} introduces the construction of the predictor and response variables, and Section \ref{sec:Model} presents \texttt{SEPNET}, a multi-task learning framework designed to jointly predict future flare and CME activity together with the probability of SEP occurrence. By learning a shared representation from these three related solar activities, the model is able to exploit common characteristics among flares, CMEs, and magnetic field measurements to improve SEP forecasting. To enhance practical capability and improve the operational forecasting of SEP events, we further refine the model architecture by considering the temporal dependency with \texttt{SEPNET-TS} and operational use with \texttt{SEPNET-O}; we present the corresponding results in Section \ref{sec:results}. The \texttt{SEPNET} model, together with \texttt{SEPNET-TS}, are initially trained using all SEP enhancements above GOES background, defined by a proton flux threshold of $10^{-6}$ pfu in the CLEAR SEP benchmark dataset, which helps mitigate class-imbalance effects and strengthens the robustness of model training. For operational deployment (\texttt{SEPNET-O}), samples labeled as operational SEP events, defined by proton flux exceeding $10$ pfu in the $>10$ MeV channel, are used as a validation set to fine-tune the classification threshold. Figure \ref{fig:Models} provides a systematic overview of the models.

\subsection{Data Preparation}\label{sec::Data_Prep}
This section details the data sources and the preprocessing procedures used to prepare the dataset for model training and evaluation. We first define the response variables and the predictor variables (input features) used in this study.

For the response variables, we use SEP records from the CLEAR SEP benchmark dataset provided by the \texttt{FetchSEP} \cite{fetchsep} python module covering the period from 3 February 1986 to 10 September 2025. This dataset includes 568 general SEP events, defined as periods when proton flux in the $>$10 MeV channel exceeds background levels (indicated with a threshold $10^{-6}$ pfu). Among these, 267 events surpass the operational threshold of 10 pfu in the $>10$ MeV proton channel, which is the criterion used by NOAA's Space Weather Prediction Center and NASA's Space Radiation Analysis Group to define a solar radiation storm or operational SEP event (see details in \url{https://www.swpc.noaa.gov/phenomena/solar-radiation-storm} and \url{https://srag.jsc.nasa.gov/spaceradiation/what/what.cfm}). Each event is characterized by the start and end times at which the $>10$ MeV proton flux crosses the respective thresholds. Note that these events are relatively rare over such an extended time period, highlighting the challenge of data sparsity in SEP forecasting studies.

The feature data sources include solar flares and CME-related features, together with SHARP parameters, from which we derive all predictors used in this study. For the flare-related features, we use the GOES flare catalog retrieved from the Heliophysics Event Knowledgebase through the SunPy Fido interface \cite{Stuart2025Sunpy,hek_dataset}, spanning from 1 September 1975 to 29 September 2025, which contains 88,492 events. For each flare, we calculate its duration (time from start to end), the rise time (time from start to peak), and the logarithm of its peak flux; and these derived quantities constitute the flare feature set. The CME-related features are obtained from the CCMC DONKI CME catalog covering the period from 3 April 2010 to 25 September 2025, totaling 7,507 events (available at \citeA{ccmc_donki}). For each CME, we extract the features of latitude, longitude, half angle, and speed, which form the CME feature set. The catalog includes both CMEs originating from active regions and non-active-region events, such as streamer blowouts, and does not further separate CMEs by type in this work. SHARP parameters, which are scalar quantities derived from full photospheric vector magnetic field magnetograms with a 12-minute cadence (see \citeA{Bobra2014SHARP} for detailed methodology), are included as well. All the SHARP parameters, i.e., LAT MIN, LON MIN, LAT MAX, LON MAX, USFLUX, MEANGAM, MEANGBT, MEANGBZ, MEANGBH, MEANJZD, TOTUSJZ, MEANALP, MEANJZH, TOTUSJH, ABSNJZH, SAVNCPP, MEANPOT, TOTPOT, MEANSHR, SHRGT45, SIZE, SIZE ACR, NACR, and NPIX are included in our study. The detailed description of these SHARP parameters, including their physical meanings and units, is provided in Table S1 of the Supporting Information. The SHARP dataset, provided by the Stanford Joint Science Operations Center (see \url{http://jsoc.stanford.edu/ajax/lookdata.html}) and accessed with the SunPy package drms \cite{Glogowski2019,Barnes2020,Stuart2025Sunpy}, ranges from 1 May 2010 to 30 September 2025 with a total of 2,632,097 records. 

\begin{figure}[hpbt]
	\centering	\includegraphics[width=1\linewidth]{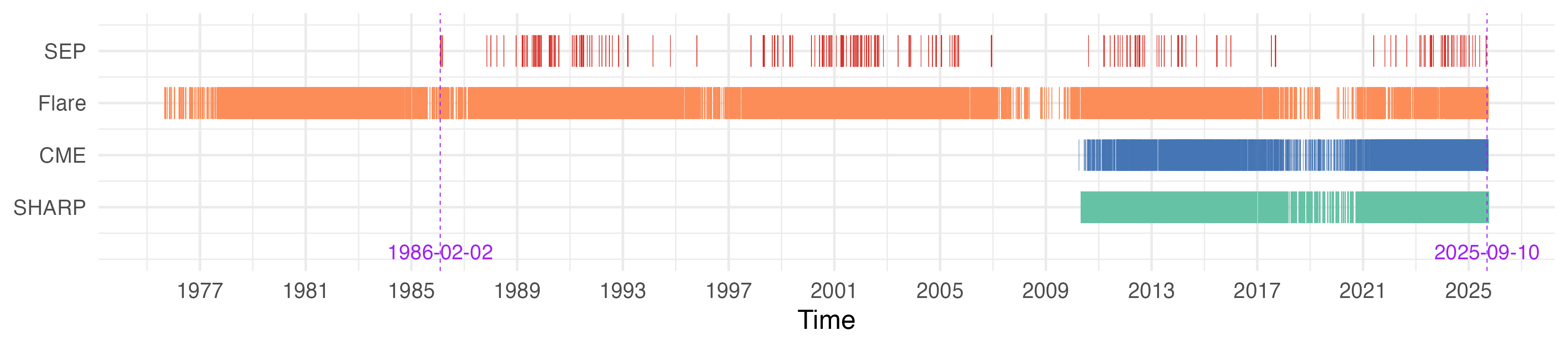}
	\caption{Visualization of the timeline for SEP (proton flux $>$ 10 pfu in the $>$ 10 MeV channel), flare, CME, and SHARP records used in this study. For each data source, only records occurring between 24 hours before the first SEP event search time and the minimum of the latest recorded times across all sources are included. Vertical purple dashed lines indicate the selected time period.}
	\label{fig:placeholder}
\end{figure}

For all feature sources (flare, CME, and SHARP), we aligned the time range used for predictor construction with the temporal coverage of the CLEAR SEP benchmark dataset. Since the earliest SEP event in the benchmark starts on 3 February 1986, the input data stream is initialized 24 hours earlier, on 2 February 1986, to enable construction of the first 24-hour predictor window. This temporal alignment was chosen to maintain consistency with the SEP records and was not intended to exclude quiet-Sun periods or to artificially rebalance the dataset. The full set of SEP, flare, CME, and SHARP events utilized in this study is visualized in Figure \ref{fig:placeholder}. We describe how all these sources of information are processed to create predictors for SEPs in Section \ref{sec::DataPreprocessing}. 

The SHARP dataset provides condensed measurements at a 12-minute cadence. However, several features contain missing values except for USFLUX, TOTUSJZ, TOTUSJH, ABSNJZH, SAVNCPP, and TOTPOT during our download periods. For the remaining SHARP features with partially missing entries, we applied a $k$-nearest-neighbors imputation approach with $k=10$ as a pragmatic preprocessing step to retain usable magnetic-field samples. Missing values were estimated as weighted averages of the corresponding feature values from nearest neighboring samples identified in feature space. Since SEP forecasting in this study relies on summary statistics derived within each 24-hour predictor window, the missing value treatment is intended to support stable feature construction and maintain sufficient samples for downstream analysis. This approach is also compatible with real-time forecasting, as it uses only past and current information, whereas interpolation-based methods often require future values. A systematic evaluation of different choices of $k$ and alternative imputation schemes will be examined in future work.

\subsubsection{Data Preprocessing}\label{sec::DataPreprocessing}
In this study, we construct the training and testing datasets using a rolling-window approach with fixed, non-overlapping 24-hour windows spanning the full temporal coverage of the CLEAR SEP benchmark dataset. For each 24-hour predictor window, we collect all flares, CMEs, and SHARP records whose timestamps fall within that interval. We then aggregate these records by computing summary statistics for each feature, specifically the minimum, maximum, and mean values across all records within the window. This aggregation yields a single feature vector per predictor window. 

\begin{figure}
	\centering
\includegraphics[width=1\textwidth]{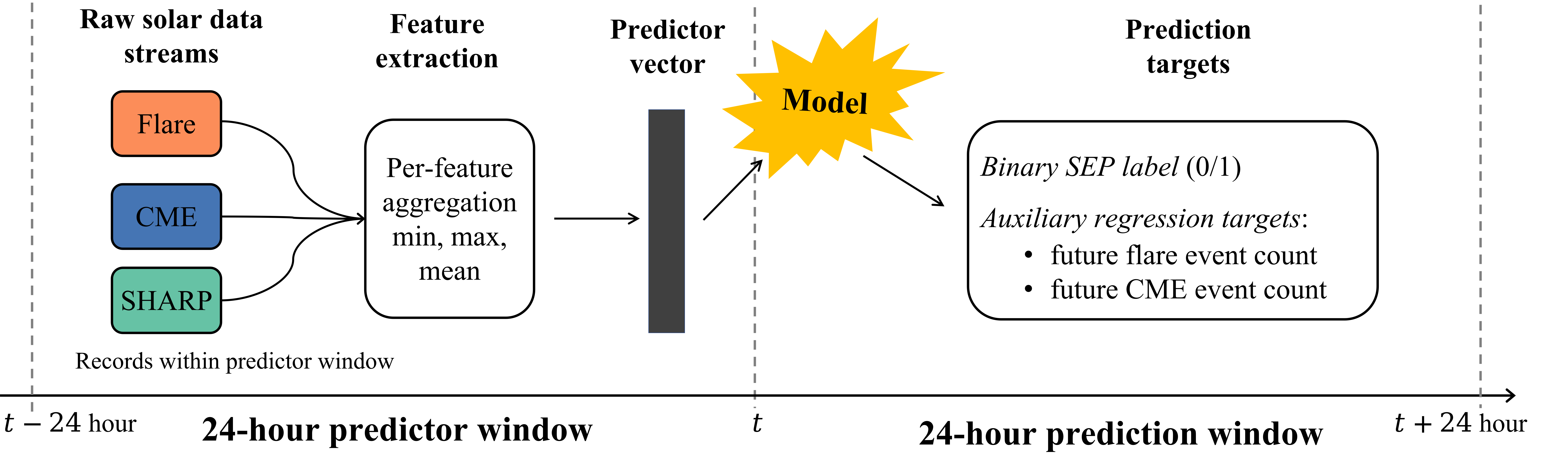}
	\caption{Schematic illustration of the data preprocessing and sample-construction procedure. Solar flares, CMEs, and SHARP records falling within a 24-hour predictor window are aggregated into summary statistics (minimum, maximum, and mean) to form a single predictor vector. The corresponding label is determined by whether an SEP event occurs in the immediately following, contiguous 24-hour prediction window. For the multi-task setting, the numbers of flare and CME events in the prediction window are also used as auxiliary targets.}
	\label{fig:preprocess}
\end{figure}

The response label for a predictor window above is an indicator of whether an SEP event occurs in the subsequent 24-hour prediction window, immediately following the predictor window (i.e., a non-overlapping lead time of 24 hours). If no SEP occurs during that prediction window, the sample is labeled as a non-SEP event (negative); otherwise, it is labeled as an SEP event (positive). This setup enables the model to forecast SEP occurrence in the next 24 hours using aggregated solar activity information from the preceding 24 hours. Summary statistics calculated from a 24-hour predictor window form a single timestep sequence for \texttt{SEPNET-TS}. The LSTM transformer layers process these sequences to learn non-linear feature interactions within each window's summary statistics. A schematic overview of the data preprocessing and sample-construction procedure is shown in Figure \ref{fig:preprocess}.

By utilizing these summary statistics, we bypass the need to explicitly map specific flares, CMEs, or SHARP parameters to SEP events, which allows us to capture more comprehensive information about the solar environment without the risk of discarding data that cannot be strictly matched. For our multi-task learning model, we also predict the number of flare and CME events in the 24-hour forecast window and record these counts for subsequent analysis. After removing predictor windows in which all flare, CME, and SHARP inputs are simultaneously missing, the dataset comprises 11,773 predictor windows; 3,537 of these windows are labeled as positive for future general SEP occurrence and are used in training and evaluation for \texttt{SEPNET} and \texttt{SEPNET-TS}. Since a single SEP event can persist for longer than 24 hours, one physical SEP event may contribute positive labels to multiple consecutive prediction windows; therefore, the number of positive predictor windows is larger than the number of distinct SEP events in the CLEAR SEP benchmark dataset. Within this same set of predictor windows, 1,726 correspond to future operational SEP occurrences and are used for threshold calibration and evaluation of \texttt{SEPNET-O}.

Since we will evaluate model performance on the SEPVAL dataset, we use the designated periods specified in the SEPVAL dataset when constructing the testing set. The SEPVAL dataset is available on Zenodo \cite{whitman2025SEPVAL} with supplementary resources provided
at \url{https://ccmc.gsfc.nasa.gov/community-workshops/ccmc-sepval-2023/}. The SEPVAL dataset comprises 33 SEP and 30 non-SEP events from 2011 to 2023. Notably, most non-SEP events have strong flares associated with them, adding further complexity to the challenge of distinguishing SEP from non-SEP intervals. For the SEPVAL evaluation, we designate as test samples those 24-hour predictor windows whose subsequent 24-hour prediction window includes the documented onset time of each SEPVAL SEP event, while all remaining predictor windows are utilized for training. To avoid unequal testing sample sizes and to retain physically meaningful quiet intervals when only flare or CME features are used across different feature selection scenarios, especially when no flare or CME is recorded in a given 24-hour window, we do not discard these samples or treat them as generic missing data. Instead, flare- and CME- related features are set to zero, and the logarithm of the flare peak flux is fixed to $-10$ to represent background or null activity \cite{Winter2015SW001170}.

After splitting the data into training and testing samples, all input features were normalized to the range $[0,1]$ using min-max scaling defined by $x^\prime=(x-x_{\text{min}})/(x_{\text{max}}-x_{\text{min}})$, where $x_{\text{min}}$ and $x_{\text{max}}$ are the minimum and maximum values computed over the training set \cite{hastie2009elements}. Information from the testing set is not used in the normalization step to avoid information leak.

\subsubsection{Feature Selection}
\label{sec:featureselection}

Given shared characteristics among SHARP parameters, CME properties, and solar flare characteristics (e.g., \citeA{Liu2017Predicting} and \citeA{Jiao2020Flare}), we systematically investigated combinations of these feature groups to optimize predictive performance and mitigate overfitting. Since the flare, CME, and SHARP data sources cover different time periods, the number of usable predictor windows varies with the selected feature subset. Table \ref{tab:size} summarizes the sample sizes for different feature subsets used in this study. Notably, using flare data alone yields the largest sample size due to its longest temporal coverage; however, subsequent results show that, despite the larger data volume, models trained solely on flare data perform sub-optimally. For each candidate feature subset, the multi-task learning model was reinitialized and trained epoch-wise to ensure consistent evaluation.
\begin{table}[htbp]
	\small
	\centering
	\begin{threeparttable}
		\caption{Number of predictor windows available for each feature subset. }
		\label{tab:size}
		\begin{tabular}{lccccccc}
			\toprule
			& F & C & F+C & S & S+F & S + C & S+F+C \\ \midrule
			General SEP & 3334 & 997 & 829 & 1260 & 1059 & 993 & 827 \\
            Operational SEP & 1635 & 455 & 376 & 585 & 494 & 455 & 376 \\
			Non-SEP & 7071 & 2232 & 1592 & 3550 & 2532 & 2042 & 1549 \\
			\bottomrule
		\end{tabular}
		\begin{tablenotes}
			\footnotesize
			\item \textbf{Abbreviations:} F = flare-related features, S = SHARP parameters, C = CME-related features. General SEP: $>$ $10$ MeV proton flux $>$ $10^{-6}$ pfu. Operational SEP: $>$ $10$ MeV proton flux $>$ $10$ pfu.
		\end{tablenotes}
	\end{threeparttable}
\end{table}

% \begin{table}[htbp]
% 	\small
% 	\centering
% 	\begin{threeparttable}
% 		\caption{Data size across different feature sets. }
% 		\label{tab:size}
% 		\begin{tabular}{lccccccc}
% 			\toprule
% 			& F & C & F+C & S & S+F & S + C & S+F+C \\ \midrule
% 			SEP & 625 & 167 & 148 & 198 & 178 & 167 & 148 \\
% 			Non-SEP & 9780 & 3062 & 2273 & 4612 & 3413 & 2868 & 2228 \\
% 			\bottomrule
% 		\end{tabular}
% 		\begin{tablenotes}
% 			\small
% 			\item \textbf{Features abbreviations:} F = flare-related features, S = SHARP parameters, C = CME-related features. 
% 		\end{tablenotes}
% 	\end{threeparttable}
% \end{table}

% \subsubsection{Handling Data Imbalance}
% Due to the rarity of SEP events compared to non-events, the dataset exhibits significant class imbalance, which can bias model predictions and reduce the reliability of the classifier. To mitigate this, a hybrid SMOTE-ENN resampling strategy \cite{BatistaSMoteENN2004} is applied exclusively to the training data after feature normalizing. The normalized input features are concatenated with the log-transformed flare and CME count targets, and SMOTE-ENN is used to generate synthetic minority SEP samples while removing noisy majority with a sampling ratio of 0.5. The resampled data are then separated back into features and regression targets and used for model training, while the original unresampled normalized training data are retained as the validation set, and the untouched normalized test set is reserved for final evaluation. 

\subsection{Model Architecture}\label{sec:Model}
We developed a multi-task neural network model, \texttt{SEPNET}, to capture the complex relationship among solar flares, CMEs, and SHARP parameters in relation to SEP occurrences. By simultaneously learning to predict flare and CME features along with SEP events in a multi-task framework, \texttt{SEPNET} leverages the shared information across these related solar phenomena, with the primary goal of SEP forecasting. This integrated approach enables the model to adaptively utilize predictive signals from flare and CME dynamics to improve the accuracy of SEP event forecasts within the next 24 hours.

\begin{figure}
	\centering
\includegraphics[width=0.9\textwidth]{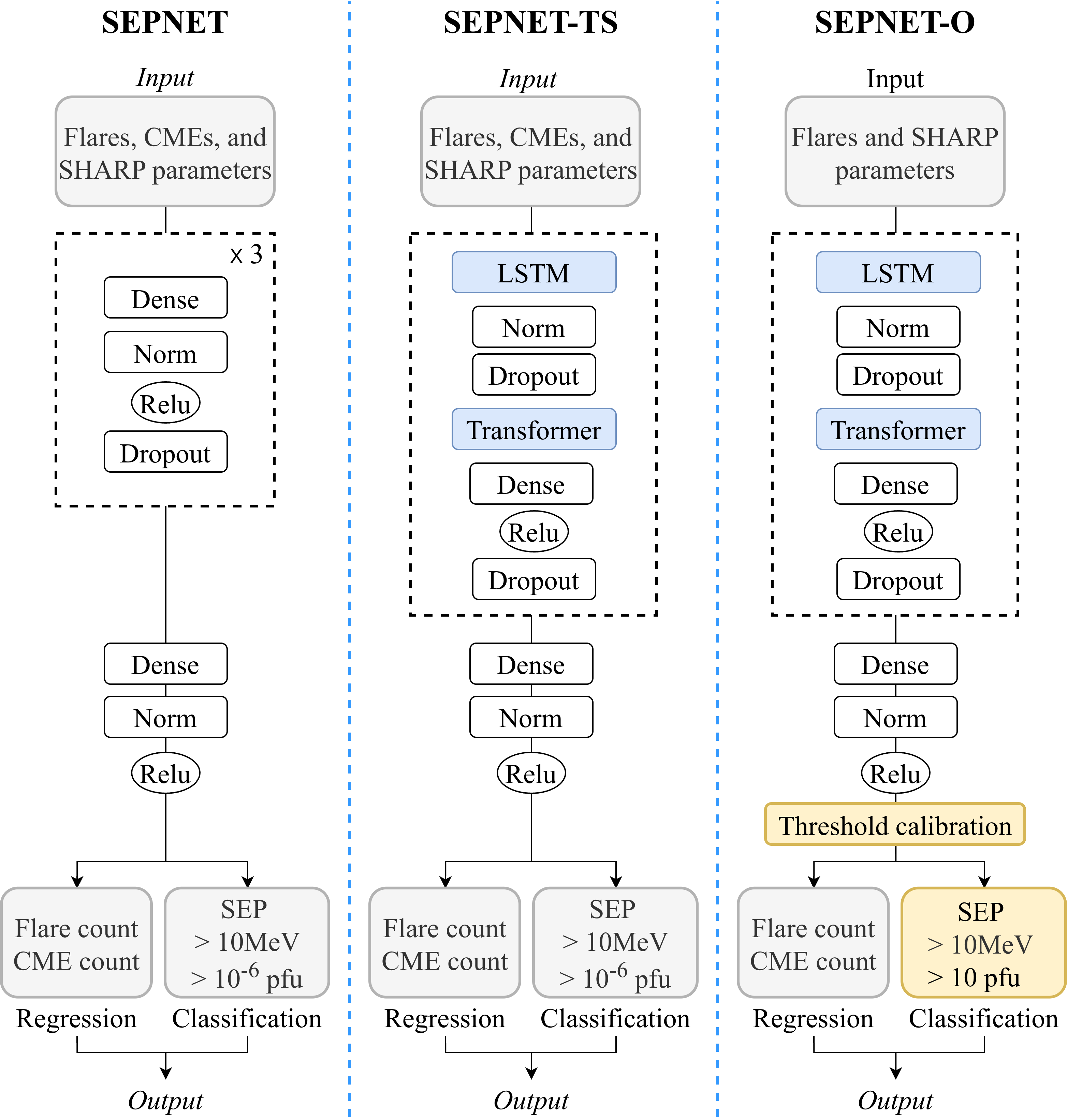}
	\caption{Diagram illustrating the architectures of the proposed multi-task learning models. \textbf{Left}: \texttt{SEPNET}, composed of shared feed-forward layers with layer normalization, ReLU activations, and dropout, followed by regression and classification heads for predicting flare/CME counts and SEP event probability. \textbf{Middle}: \texttt{SEPNET-TS}, an updated version introducing sequential processing via a unidirectional LSTM and transformer encoder before multi-task prediction. \textbf{Right}: \texttt{SEPNET-O}, an version of \texttt{SEPNET} for real time operational SEP prediction.}
	\label{fig:Models}
\end{figure}

The flowchart of the \texttt{SEPNET} model is depicted in the left panel of Figure \ref{fig:Models}. For each sample, the input consists of a set of min-max normalized features derived from solar flare, CME, and SHARP magnetic field data. These features are processed through three shared fully connected (dense) layers with gradually reduced feature dimensionality (from $256$ to $128$, $64$, and $16$), which encourages the formation of efficient, compressed representations and facilitating hierarchical feature extraction \cite{Wu2020Understanding,wang2024timit}. Each dense layer is followed by layer normalization, which stabilizes training, mitigates issues arising from varying input scales, and is particularly beneficial for deep multilayer perceptrons. ReLU activation functions, ReLU$(x)=\max(x,0)$, introduce nonlinearity \cite{zou2020gradient}. Dropout is applied after activation to prevent coadaptation among neurons and reduce the risk of overfitting. The staged compression helps filter noise and focuses network capacity, ensuring the final shared representation remains suitably compact for both tasks while balancing model complexity and computational efficiency. The shared embedding is then fed into two distinct output heads to implement multi-task learning: a regression head that predicts the counts of future flare and CME events, and a classification head that outputs the predicted probability for the occurrence of a future SEP event. This architectural choice leverages feature sharing to boost learning efficiency while allowing task-specific prediction at the output layer \cite{MultitaskAnders2024}.

The updated model \texttt{SEPNET-TS} replaces \texttt{SEPNET}'s dense layers with a more expressive LSTM-transformer backbone for sophisticated feature transformation, illustrated in the middle panel of Figure \ref{fig:Models}. Each 24-hour aggregated predictor window is formatted as a predictor vector and processed through: (1) a unidirectional LSTM layer with hidden dimension 64 to capture sequential dependencies among the summary statistics derived from feature inputs within each window; (2) layer normalization and dropout; (3) a transformer encoder with 1 layer and 4 attention heads, and a final shared representation dimension of 16 for hierarchical attention-based feature interactions; and (4) final timestep output fed to separate regression (flare/CME counts) and classification (SEP probability) linear heads. This architecture leverages sequential layers for rich non-linear feature engineering from the summary statistics while maintaining computational efficiency of single-window processing \cite{cao2024advanced,ZHANG2025110732}.

\subsubsection{Loss Function}
Models are trained with a joint loss function that combines mean squared error (MSE) for regression and binary cross-entropy with sigmoid activation (BCEWithLogitsLoss) for classification. The future flare and CME counts are highly concentrated near zero and exhibit right-skewed distributions with occasional large values. To reduce the influent of this skewness during model training, a logarithmic transformation was applied after incrementing all count values by 1 to avoid numerical underflow. The distributions for these two targets are visualized in Figure S1 of the Supporting Information.

For the SEP classification task, we use a combined objective consisting of binary
cross-entropy and an additional focal-loss term \cite{lin2017focal}. The binary cross-entropy term provides a
stable baseline classification objective, while focal loss places additional
emphasis on hard and minority-class examples under severe class imbalance.
Empirically, using BCEWithLogitsLoss alone with a weighted sampler led to a
relatively high false-positive rate, whereas the combined formulation performed
more reliably than focal loss alone and helped reduce false alarms. However, alternative loss formulations may further improve predictive performance and should be evaluated more systematically in future work.

Formally, let $\hat{y}_{ _\text{Flare},t}$ and $\hat{y}_{ _\text{CME},t}$ denote the predicted future event counts for flares and CMEs at time $t$, respectively, with $y_{ _\text{Flare},t}$, $y_{ _\text{CME},t}$ representing the corresponding ground truth values. Let $\hat{y}_{_\text{SEP},t} \in [0,1]$ denote the estimated probability of a general SEP event, where the true binary label is defined as $y_{_\text{SEP},t}\in \{0,1\}$, where $y_{_\text{SEP},t}=1$ indicates the occurrence of an SEP event within the subsequent 24 hours and $0$ otherwise. The overall loss function is defined as $\mathcal{L}=\mathcal{L}_{\text{MSE}} + \mathcal{L}_{\text{BCEWithLogits}}+ \lambda \mathcal{L}_{\text{Focal}},$ where \begin{align*}
\mathcal{L}_{\text{MSE}} =& \frac{1}{2N}\sum_{t=1}^N \left[
 \left(\log(\hat{y}_{ _\text{Flare},t} {+} 1) - \log(y_{_\text{Flare},t} {+} 1) \right)^2 \right. \\
 &\qquad \qquad +\left. \left(\log(\hat{y}_{ _\text{CME},t} {+} 1) - \log(y_{ _\text{CME},t} {+} 1)\right)^2 \right], \\
\mathcal{L}_{\text{BCEWithLogits}} =& \frac{1}{N}\sum_{t=1}^N \left[ 
 -y_{_\text{SEP},t}\log \hat{y}_{_\text{SEP},t} - (1{-}y_{_\text{SEP},t})\log(1{-}\hat{y}_{_\text{SEP},t}) \right], \\
\mathcal{L}_{\text{Focal}} =& \frac{1}{N}\sum_{t=1}^N \alpha\,(1{-}\hat{y}_{_\text{SEP},t})^\gamma \log \hat{y}_{_\text{SEP},t}.
\end{align*} 
Here, $N$ is the total number of samples. The Focal loss hyperparameters are set by default as $\alpha=0.25$, a balancing factor that weights the minority class more heavily, and $\gamma=2$, a focusing parameter that adjusts the degree to which easy examples are down-weighted. As $\gamma \rightarrow 0$, the Focal loss converges to the standard cross-entropy loss. The scalar weight $\lambda$ balances the contribution of the Focal loss relative to the other components and is set to $10$ based on the relative scale of the losses in our experiments.

\subsubsection{Evaluation Metrics}
Model performance was evaluated on the test set using criteria routinely adopted in the space weather community, including both threshold-agnostic and event-based confusion-matrix-derived metrics. Specifically, the analysis includes accuracy (ACC), area under the receiver operating characteristic curve (AUC; \citeA{muschelli2020roc}), F1 score (threat score; \citeA{lipton2014f1}), probability of detection (POD, recall, hit rate; \citeA{wehling2011probability}), false positive rate (FPR), false alarm ratio (FAR; \citeA{macmillan1985detection}), true skill score (TSS; \citeA{doswell1990summary}), and Heidke skill score (HSS). The emphasis is placed on F1, POD, TSS and HSS, which capture core operational priorities such as sensitivity and event capture skill \cite{Leka2019}. While ACC and AUC are standard for general classification, these metrics can mask important shortcomings in imbalanced-event scenarios, making confusion-matrix-based measures essential in SEP forecasting, where both false positives (FP) and missed detections (FN) have significant operational implications. For statistical robustness, all metrics are aggregated across 50 random seeds to reliably quantify model performance with reduced variability. Detailed definitions and formulas for these evaluation criteria are provided in the Supporting Information.

\subsubsection{Hyperparameter Selection and Optimization}\label{sec:hyperparameter}

In this study, hyperparameter selection is performed using the Optuna \cite{akiba2019optuna}, an automated optimization framework designed for efficient exploration and pruning of parameter combinations. Optuna uses a define-by-run approach, enabling dynamic specification of search spaces and flexible experiment definition, which is particularly advantageous for neural network architectures. 

Specifically, the following hyperparameters were tuned: the learning rate, which is sampled log-uniformly within $[10^{-5}, 10^{-3}]$, dictating the magnitude of parameter updates during Adam optimization; the weight decay, influencing the $L_2$ regularization strength to alleviate overfitting; the dropout probability, uniformly sampled from $[0.1,0.5]$, which further mitigates overfitting within hidden layers; and the batch size, which balances training stability with computational efficiency. The focal-loss parameters $\alpha=0.25$ and $\gamma=2$ were fixed, and the weighting factor $\lambda=10$ was chosen empirically.

The objective function is defined on model training and returns the average training loss over repeated runs to account for stochasticity. Optuna's optimization algorithm~\cite{akiba2019optuna} systematically evaluates these trials and prunes unpromising candidates early, focusing resources on promising configurations. The final hyperparameter configuration corresponds to the lowest observed training loss.

Training incorporates gradient clipping, which constrains the norm of model gradients to improve stability and prevent divergence. A learning rate scheduler is used, reducing the learning rate when validation loss plateaus during training to improve convergence near the optimum. Additionally, early stopping halts training when no improvement in training loss is observed for an extended period, further reducing overfitting and computational cost. The final selected hyperparameter values for the \texttt{SEPNET} and \texttt{SEPNET-TS} models across different experiment settings are summarized in Table S6 of the Supporting Information. 

\section{Results}\label{sec:results}
We conduct a comprehensive evaluation of SEP event prediction models, leveraging both advanced machine learning architectures and classical methods across different testing scenarios and feature combinations.

\subsection{SEPVAL Dataset Evaluation}
\begin{figure}[hpbt]
	\centering
	\includegraphics[width=1\textwidth]{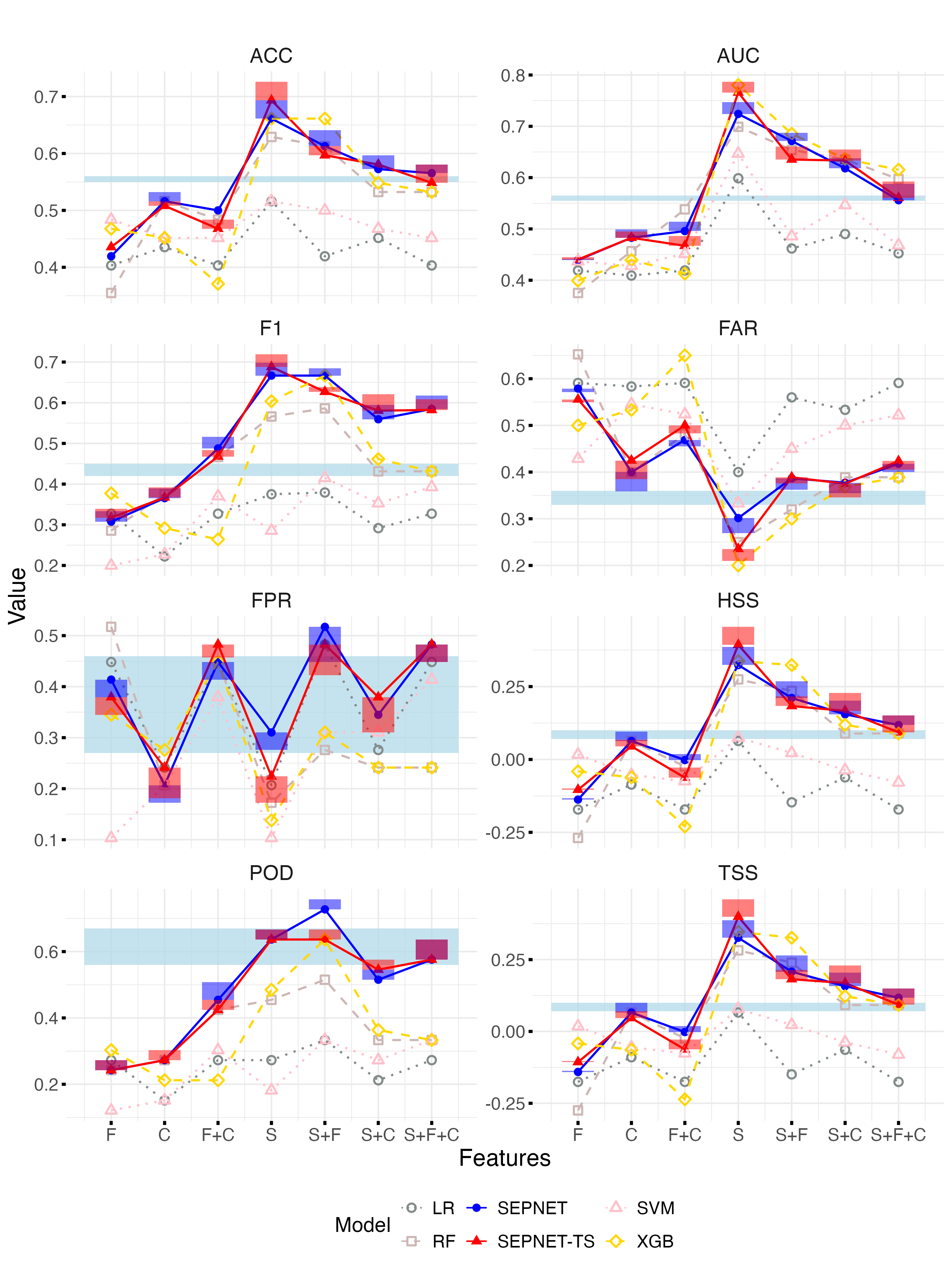}
	\caption{
		Performance metrics for SEPVAL prediction models, showing the median and target quantile values across different feature sets and model architectures. The shaded light blue region represents the median and target quantile achieved by state-of-the-art pre-eruption models. Feature set abbreviations: F = flare-related features; S = SHARP parameters; C = CME-related features. Performance metric abbreviations: ACC = accuracy; AUC = area under the curve; FPR = false positive rate; F1 = F1 score; POD = probability of detection; FAR = false alarm ratio; TSS = true skill score; HSS = Heidke skill score. Model abbreviations: LR = logistic regression with elastic net regularization; SVM = support vector machines; RF = random forests; XGB = extreme gradient boosting.}
	\label{fig:SEPVAL1}
\end{figure}

In this section, we focus on the SEPVAL test dataset to evaluate the performance of our \texttt{SEPNET} model and its updated version, \texttt{SEPNET-TS}, which processes each 24-hour aggregated predictor window as a sequence through LSTM transformer layers to learn rich non-linear feature representation from the summary statistics. Various combinations of input features were tested, and the results are detailed in Table S2 of the Supporting Information and visualized in Figure \ref{fig:SEPVAL1}. The results demonstrate that models employing SHARP parameters, either alone or combined with flare-related features, yield better predictive performance. Conversely, models relying exclusively on flare and CME features exhibit lower skill, with some metrics, such as TSS and HSS, falling below zero, indicating limited capability to reliably forecast SEP events within the subsequent 24 hours.

For a rigorous comparison with the state-of-the-art pre-eruptive models (denoted as SoA) on SEPVAL \cite{Whitman2026}, we performed 50 independent runs for each configuration to derive the median and $75$th percentile (target quantile) metrics. Our models incorporating SHARP parameters generally match or outperform the SoA benchmarks in terms of standard evaluation metrics such as ACC, AUC, F1, POD, TSS, and HSS. Notably, \texttt{SEPNET} and \texttt{SEPNET-TS} models with SHARP features substantially surpass the SoA models, underscoring superior capability for detecting SEP event occurrence. However, the relatively high false alarm rate highlights the ongoing challenges in achieving high specificity.

In addition, we benchmark classical machine learning techniques, including logistic regression with elastic net regularization (LR), support vector machines (SVM), random forests (RF), and extreme gradient boosting (XGB) for SEP classification tasks on the SEPVAL dataset; detailed results are provided in Table S3 of the Supporting Information. Among these, the XGB model attained the best overall performance, particularly when trained on SHARP parameters alone or in combination with flare-related features. Nevertheless, these classical single-task approaches consistently fall short of the predictive skill delivered by our proposed \texttt{SEPNET} architectures. To further illustrate the contribution of the multi-task formulation, we also compared \texttt{SEPNET-TS} with a matched single-task variant that used the same backbone architecture but omitted the regression head and was trained solely for SEP classification. The corresponding results, presented in Table S2 of the Supporting Information, show that the multi-task model achieved improved SEP forecasting performance, suggesting that the auxiliary flare- and CME-related tasks help the model learn more informative shared representations for the primary SEP prediction task.

\subsection{Stratified Random Split Evaluation}
Given the limited number of SEP events and non-events in SEPVAL, we run the evaluations using a stratified random split for the full CLEAR dataset. More precisely, we use a random stratified split, allocating 20\% of the dataset to testing and the remaining 80\% to training. This split is repeated five times with different random seeds to ensure a robust assessment, and median metric values across the replicates are reported. This approach reaffirmed the superior performance of \texttt{SEPNET} models in terms of F1 score, POD, and skill scores (TSS and HSS), as presented in Figure \ref{fig:Scenario2} and detailed in Table S4 of the Supporting Information.

The regression results for the auxiliary flare-count and CME-count prediction tasks are reported in Table \ref{tab:S2-reg}. Overall, the test performance is broadly consistent with the training performance, exhibiting slightly larger errors, as expected. Among the evaluated configurations, \texttt{SEPNET-TS} incorporating both SHARP parameters and flare-related features attains the best overall performance on the test sets. The corresponding training and inference computational costs, aggregated over the five independent random stratified splits, are presented in Table \ref{tab:S2-cost}. Although the regression errors remain non-negligible, these results indicate that the auxiliary tasks capture physically meaningful variability and that the inference costs of \texttt{SEPNET} and \texttt{SEPNET-TS} remain computationally feasible for near-real-time forecasting applications. Further refinement and optimization of the auxiliary regression heads therefore constitute an important direction for future research.

\begin{figure}[hpbt]
	\centering
	\includegraphics[width=1\textwidth]{Scenario2.png}
	\caption{Performance metrics on the 20\% testing set for different feature sets and models, targeting classification of general SEP events. Results for each criterion are the median values across five independent random stratified data splits. Feature set abbreviations: F = flare-related features; S = SHARP parameters; C = CME-related features. Performance metric abbreviations: ACC = accuracy; AUC = area under the curve; FPR = false positive rate; F1 = F1 score; POD = probability of detection; FAR = false alarm ratio; TSS = true skill score; HSS = Heidke skill score. Model abbreviations: LR = logistic regression with elastic net regularization; SVM = support vector machines; RF = random forests; XGB = extreme gradient boosting.}
	\label{fig:Scenario2}
\end{figure}

\begin{table}[htbp]
	\small
	\centering
	\begin{threeparttable}
		\caption{Regression performance for future flare-count and CME-count prediction under stratified random split evaluation. Results are the median values across five independent random stratified data splits.}
		\label{tab:S2-reg}
		\begin{tabular}{llccccccc}
			\toprule
			\multirow{2}{*}{Features} & \multirow{2}{*}{Model} & \multirow{2}{*}{Split} & \multicolumn{3}{c}{Flare count} & \multicolumn{3}{c}{CME count} \\ \cmidrule(lr){4-6} \cmidrule(lr){7-9}
            &  & & RMSE & MAE & $R^2$  & RMSE & MAE & $R^2$ \\
                        \midrule
			\rowcolor[HTML]{EFEFEF}
			S & \texttt{SEPNET} & Training set & 3.5059 & 2.3218 & 0.3964 & 1.4695 & 1.0231 & 0.3102 \\
			\rowcolor[HTML]{EFEFEF}
			      & \texttt{SEPNET} & Testing set  & 4.0293 & 2.8038 & 0.2000 & 1.5783 & 1.1401 & 0.1324 \\
                  \rowcolor[HTML]{EFEFEF}
			      & \texttt{SEPNET-TS} & Training set & 3.4799 & 2.2810 & 0.4054 & 1.4746 & 1.0184 & 0.3455 \\
                  \rowcolor[HTML]{EFEFEF}
			      & \texttt{SEPNET-TS} & Testing set  & 3.9966 & 2.7591 & 0.2130 & 1.5404 & 1.1078 & 0.1735 \\
			S+F & \texttt{SEPNET} & Training set & 3.6683 & 2.5594 & 0.3359 & 1.4526 & 1.0317 & 0.3228 \\
			              & \texttt{SEPNET} & Testing set & 3.9164 & 2.8250 & 0.2199 & 1.4864 & 1.1919 & 0.2236 \\
			              & \texttt{SEPNET-TS} & Training set & 3.4688 & 2.3811 & 0.4456 & 1.4080 & 0.9935 & 0.3524 \\
			              & \texttt{SEPNET-TS} & Testing set & 3.8912 & 2.8985 & 0.2508 & 1.4806 & 1.0866 & 0.2547 \\
			\bottomrule
		\end{tabular}
		\begin{tablenotes}
			\footnotesize
			\item \textbf{Notes:} Feature abbreviations: S = SHARP parameters; S+F = SHARP parameters with flare-related features. Performance metric abbreviation: RMSE = root mean square error; MAE = mean absolute error; $R^2$ = coefficient of determination. Lower RMSE and MAE indicate smaller prediction error, while larger $R^2$ indicates better explained variance.
		\end{tablenotes}
	\end{threeparttable}
\end{table}

\begin{table}[htbp]
	\small
	\centering
	\begin{threeparttable}
		\caption{Training and inference time for each repeat under stratified random split evaluation. Results are median values across five independent random stratified data splits.}
		\label{tab:S2-cost}
		\begin{tabular}{llcc}
			\toprule
			Features & Model & Training time (sec) & Inference time (sec) \\ \midrule
			\rowcolor[HTML]{EFEFEF}
			S & \texttt{SEPNET}    & 58.7514 & 0.0035 \\
			\rowcolor[HTML]{EFEFEF}
			  & \texttt{SEPNET-TS} & 192.4531 & 0.0110 \\
              	\rowcolor[HTML]{EFEFEF}
			  & LR                 & 0.4396 & 0.0001 \\
              	\rowcolor[HTML]{EFEFEF}
			  & SVM                & 1.3041 & 0.1077 \\
              	\rowcolor[HTML]{EFEFEF}
			  & RF                 & 1.2935 & 0.0074 \\
              	\rowcolor[HTML]{EFEFEF}
			  & XGB                & 0.1801 & 0.0005 \\
			S+F & \texttt{SEPNET}    & 22.1870 & 0.0031 \\
			    & \texttt{SEPNET-TS} & 73.6998 & 0.0080 \\
			    & LR                 & 0.3406 & 0.0001 \\
			    & SVM                & 0.8509 & 0.0743 \\
			    & RF                 & 1.0037 & 0.0061 \\
			    & XGB                & 0.2206 & 0.0007 \\
			\bottomrule
		\end{tabular}
		\begin{tablenotes}
			\footnotesize
			\item \textbf{Notes:} Features abbreviations: S = SHARP parameters; S+F = SHARP parameters with flare-related features. Model abbreviations: LR = logistic regression with elastic net penalty; SVM = support vector machine; RF = random forest; XGB = extreme gradient boosting. 
		\end{tablenotes}
	\end{threeparttable}
\end{table}

For operational applicability, specifically forecasting SEP events with proton fluxes exceeding 10 pfu at energies $>$ 10 MeV, the model was first trained on all general SEP events in the training dataset. The decision threshold for distinguishing operational SEP events was then re-optimized by maximizing HSS, using the operational SEP-labeled training data as the validation set. In the testing phase, operational SEP events served as reference labels, and the performance of threshold-recalibrated (re-validated) models (\texttt{SEPNET-O}) was compared with that of the original \texttt{SEPNET-TS} models on the same test set.

\begin{figure}[hpbt]
	\centering
	\includegraphics[width=1\textwidth]{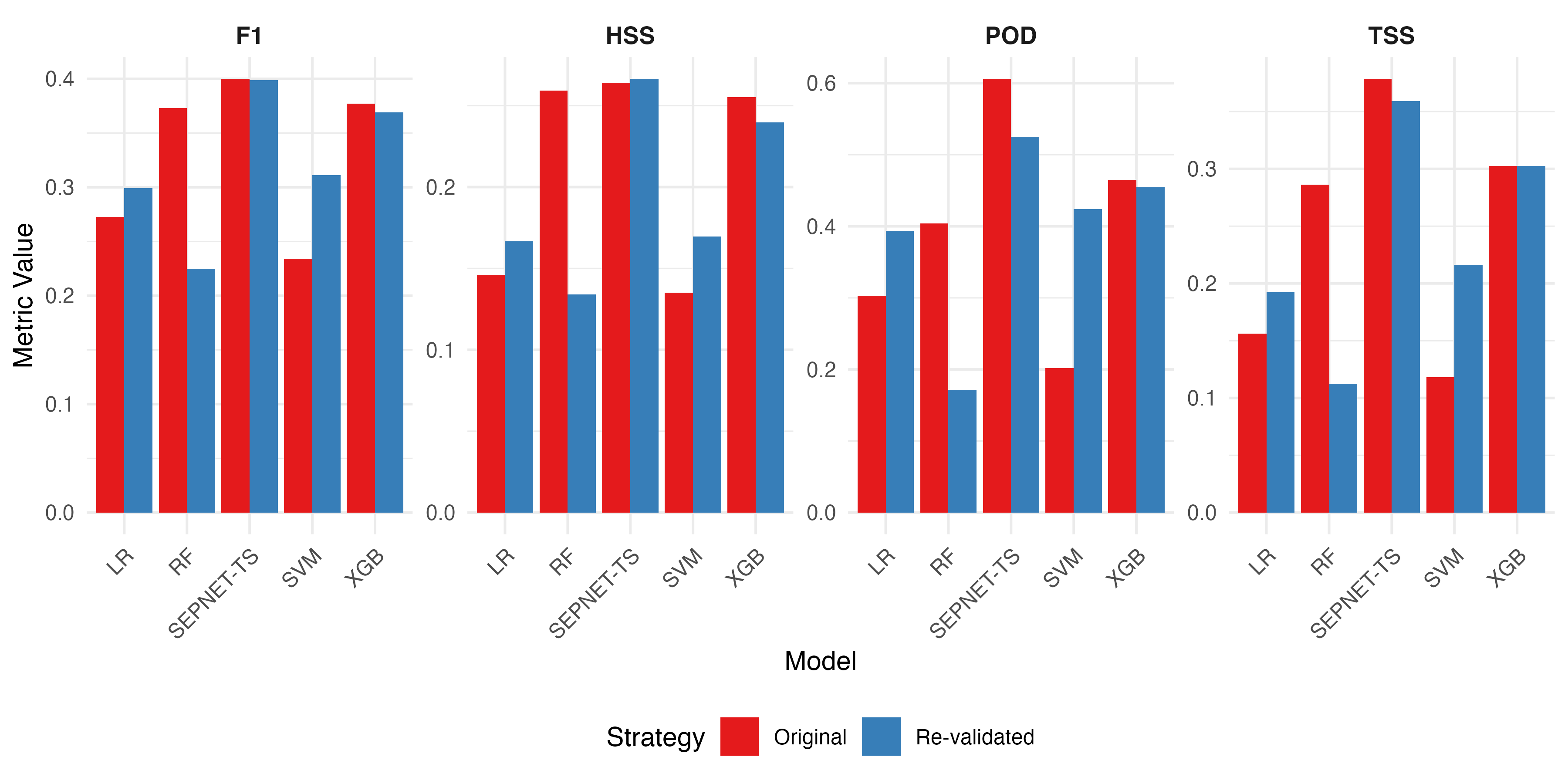}
	\caption{Performance of re-validated models (optimize the decision threshold for operational SEP event prediction) compared to original models, targeting classification of operational SEP events. Metrics are derived on the 20\% testing set using SHARP parameters with flare features, with results for each criterion being the median values across five independent random stratified data splitting. Performance metric abbreviations: F1 = F1 score; POD = probability of detection; TSS = true skill score; HSS = Heidke skill score. Model abbreviations: LR = logistic regression with elastic net regularization; SVM = support vector machines; RF = random forests; XGB = extreme gradient boosting. }
	\label{fig:Scenario2-Op}
\end{figure}

From the previous analysis, SHARP parameters combined with flare features were found to provide a suitable input set for general SEP prediction, and this feature combination is therefore adopted here for evaluating operational SEP performance. In this context, \texttt{SEPNET-TS} is compared with several classical machine learning models using the re-validation strategy, with results summarized in Table S5 of the Supporting Information and partially visualized in Figure \ref{fig:Scenario2-Op}. \texttt{SEPNET-TS} exhibits greater robustness with only modest changes in the performance criteria, while achieving comparatively higher AUC, lower FPR, and improved HSS. For operational SEP events, \texttt{SEPNET-TS} attains an accuracy close to $0.8$ and a TSS of approximately $0.36$, indicating competitive skill in distinguishing SEP events from non-SEP intervals in an operational setting. 

For completeness, the performance of \texttt{SEPNET-O} on the SEPVAL dataset is also reported in Table S2 of the Supporting Information. In this operational evaluation, \texttt{SEPNET-O} yields better performance than \texttt{SEPNET-TS} on SEPVAL, suggesting that the threshold recalibration adopted in \texttt{SEPNET-O} is beneficial when the prediction target is aligned more closely with the operational SEP definition.

\subsection{Real-time Forecasting}
In this section, we focus on the operational challenge of real-time forecasting for SEP events expected in the months following the latest entry of the CLEAR SEP benchmark dataset. All data collected up to 10 September 2025 were used for model training, which was then applied in an operational setting. For evaluation, we use the most recent flare observational features, combined with SHARP parameters, spanning from 23 October to 15 November 2025. It is important to note, however, that the SHARP parameters available in near-real-time differ from the definitive HARP data used during training (see the detailed information in \url{http://jsoc.stanford.edu/doc/data/hmi/sharp/sharp.htm}). Additionally, discrepancies in the alignment between SHARP active region designations and the corresponding flare events may result in systematic underestimation of future flare event counts. Such mismatches exemplify typical complications in real-time space weather forecasting, as highlighted in previous studies, e.g., \citeA{Bobra2015Solar}, \citeA{Leka2019}, and \citeA{chen2024solar}. 

\begin{figure}[hpbt]
	\centering
	\includegraphics[width=1\textwidth]{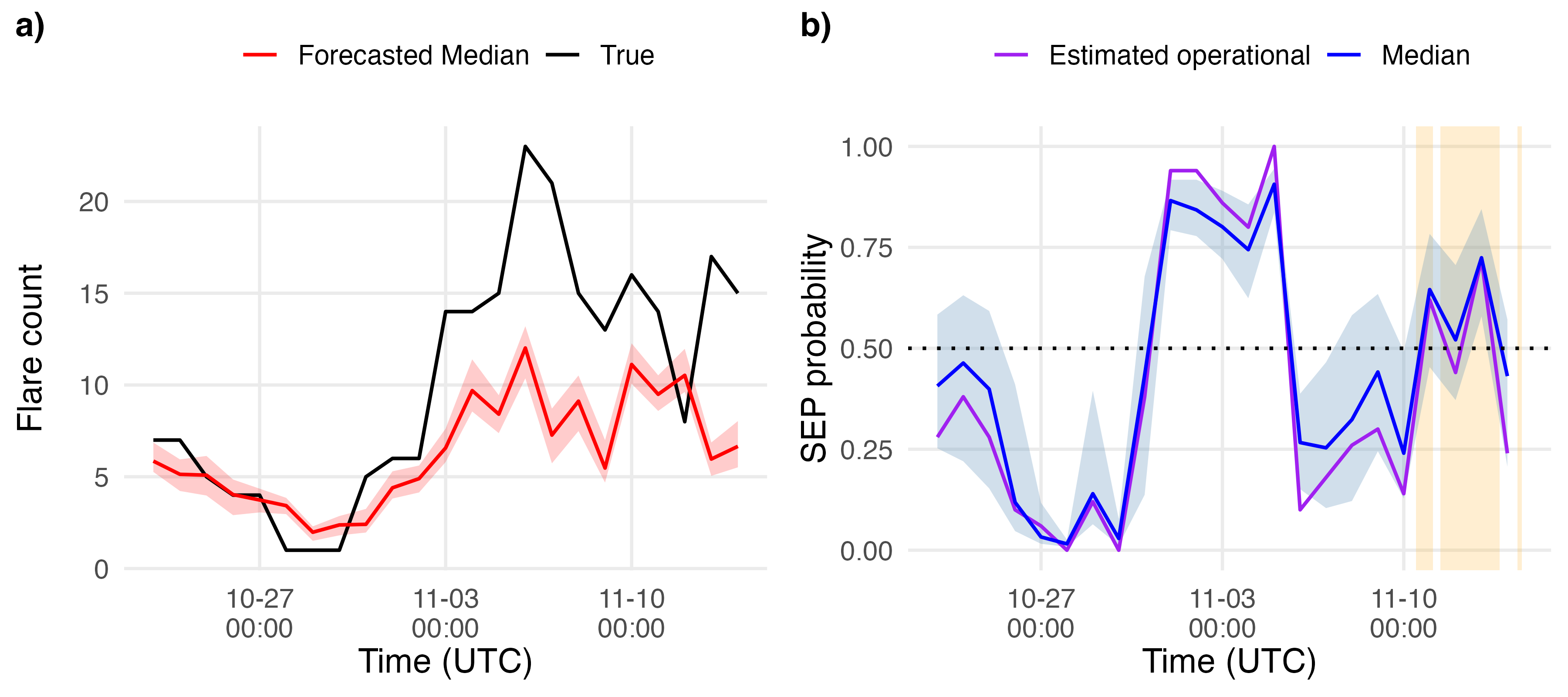}
	\caption{\texttt{SEPNET-O}'s forecasting performance for flare counts and SEP event probabilities over a recent 23-day period in November 2025. \textbf{Left panel:} The black curve indicates observed flare counts, while the red curve shows the median forecast with shaded regions representing the interquartile range ($25$th to $75$th percentiles). \textbf{Right panel:} The blue curve corresponds to the forecasted median SEP event probability, and the purple line indicates the estimated operational SEP probability based on the calibrated decision threshold, with the shaded blue region showing the interquartile range, and orange bands mark identified SEP event intervals.}
	\label{fig:RealTime}
\end{figure}

For model development, we adopted the general SEP definition encompassing all events for initial training. The model was trained using the selected hyperparameters of learning rate = $4.536$e-4, weight decay = $1.0856$e-4, dropout probability = $0.1039$, and batch size = $128$. A subsequent validation step employed the more restrictive operational SEP definition to determine an appropriate decision threshold, optimizing HSS for operational SEP event classification. The experiment was repeated 50 times to account for statistical variability. In each repetition, an optimal decision threshold was recalibrated to improve classification accuracy. The right panel of Figure \ref{fig:RealTime} therefore shows the median forecast probabilities with their 25th and 75th percentiles, as well as the estimated probability of an operational SEP event inferred from the recalibrated threshold. The resulting binary predictions align with the median of the probabilistic SEP warnings, indicating that the thresholding strategy is consistent with the underlying probability estimates.

Despite the inherent discrepancies between training and real-time datasets, \texttt{SEPNET-O} reproduces the temporal patterns of future flare and CME occurrences reasonably well and issues SEP warning probabilities for active intervals during 10-14 November 2025 (see the event list at \url{https://sep.ccmc.gsfc.nasa.gov/events.html}), as illustrated in Figure \ref{fig:RealTime}. Consistent with earlier findings, there remains a tendency toward heightened false alarm rates, most notably around 1 November, when predicted SEP risk was elevated alongside a marked flare activity. Improving the precision of SEP warnings, particularly by reducing false positives while maintaining sensitivity, will be a priority for future model development and operational deployment. 

Beyond the retrospective evaluation described here, the framework can also be operated in a continuously updated forecasting mode using near-real-time input streams with hourly updates. In this setting, forecasts of SEP occurrence over the subsequent 24 hours are refreshed every hour. These continuously updated forecasts are displayed on the University of Michigan space weather machine learning website \url{https://mlsw.engin.umich.edu/apps/sepnet}. 

\section{Discussion}\label{sec:discussion}
This study advances space weather forecasting by demonstrating the effectiveness of multi-task learning and deep neural architectures for predicting SEP events. Our approach integrates solar flares, CMEs, and SHARP magnetic field parameters, enabling the models to capture the complex interactions intrinsic to SEP generation. Evaluation against classical binary classifiers across multiple input feature sets demonstrates that the combination of flare and SHARP magnetic features yields superior predictive performance for SEP events in the next 24 hours. The findings indicate that models incorporating SHARP parameters, either alone or in combination with flare features, achieve the highest predictive skill, as reflected in F1 scores, POD, and skill scores (TSS and HSS). 

%\add[yian]{The relative performance of the different feature groups provides physics insight into the precursor information exploited by the models. Feature sets including SHARP magnetic parameters systematically yield the best SEP prediction skill, indicating that, within the current framework, the magnetic configuration provides the most informative signal. Flare-related features further enhance performance when combined with SHARP parameters, whereas CME-only inputs contribute less in the present setup. This is likely to the limited temporal coverage of the CME catalog. Further work will incorporate additional CME databases, such as the CDAW catalog \mbox{\cite{gopalswamy2025CDAW}}. A more granular attribution study at the individual-parameter level will also be undertaken in subsequent analyses.}
The comparative performance of the feature groups also offers physical insight into the precursor information leveraged by the models. Feature sets that include SHARP magnetic parameters consistently deliver the highest SEP prediction skill, indicating that, within the present framework, the magnetic configuration carries the most informative signal. Flare-related features further improve performance when combined with SHARP parameters, whereas CME-only inputs contribute less in the current setup, likely due to the limited temporal coverage of the CME catalog. Future work will incorporate additional CME databases, such as the CDAW catalog \cite{gopalswamy2025CDAW}, and will include a more detailed attribution analysis at the level of individual parameters.

An important direction for future work is to extend the magnetic-feature input beyond the HMI era. One promising option is the merged SMARP-SHARP dataset developed by \citeA{kosovich2024TSSMARP}, which combines MDI- and HMI-based active-region magnetic parameters and extends the record back to 1996. Incorporating this dataset could increase the amount of training data available for SEP forecasting, improve coverage of solar cycle 23, and enable a broader multi-cycle evaluation of SEPNET. Adapting SEPNET to this longer magnetic record will require careful treatment of cross-instrument differences between the MDI and HMI eras.

A common challenge across all scenarios is the inherent class imbalance between SEP and non-SEP intervals, which limits POD and skill score performance. While the multi-task \texttt{SEPNET} models outperform classical machine learning methods and often match or exceed SoA empirical benchmarks, one remaining limitation is the relatively high FAR. This reflects a tendency for models to overpredict, despite an increased sensitivity, can reduce operational trust and lead to unnecessary caution. In particular, some intervals associated with strong flare eruptive activity may not ultimately produce SEP events detected at the observer \cite{Swalwell2017solar}, and a more detailed event-level analysis of these false-positive cases will be necessary to better understand and mitigate such errors. Future improvements will likely derive from further dataset extension, augmentation, and integration of additional solar wind and interplanetary environment features. More sophisticated neural architectures (e.g., meta-learning, ensemble methods) and robust augmentation techniques could also yield better generalization and reliability for operational deployment.

From an operational perspective, the demonstrated ability of our models to jointly forecast SEP occurrence and the associated flare and CME activity rates within the subsequent 24-hour window offers additional predictive nuance for space weather mitigation, which will be further assessed under real-time conditions. At the same time, the auxiliary regression results indicate that there is still substantial room to improve the prediction of future flare and CME counts. While the present work does not explicitly predict SEP peak flux or fluence, the multi-output framework could be extended to include event magnitude as an additional target, with potential utility for scheduling satellite operations, astronaut extravehicular activities, power grid reconfiguration, and aviation route planning.

\section{Conclusions}\label{sec:conclusion}
In summary, our results highlight the power and promise of modern machine learning, particularly multi-task neural networks incorporating sequential dynamics for space weather prediction. By leveraging rich, multi-source solar activity data and advanced feature integration strategies, our models deliver robust, timely forecasts of SEP events, flares, and CMEs. Continued progress will depend on expanding training datasets, incorporating new physical observables, and refining model architectures to maximize event detection sensitivity while reducing false alarms. Future extensions will focus on predicting SEP integrated flux and duration, enabling a more comprehensive forecast framework that quantifies not only event occurrence but also intensity and temporal evolution, key elements for effective space weather hazard mitigation.

\section*{Open Research}
Version 1.0 of the SEPNET code used in this study is archived in Zenodo \cite{yianyu2026SEPNET} and is developed openly at GitHub \cite{yuyianSEPNet2025}. This repository provides access to the SEPVAL benchmark dataset, the \texttt{SEPNET} model implementation, and relevant analysis scripts. Users can freely access, reproduce, and build upon the research results presented in this manuscript. The model results are displayed also at the University of Michigan's space weather machine learning website: \url{https://mlsw.engin.umich.edu/apps/sepnet}. %For additional details on open research practices, please refer to the guidelines at \url{https://www.agu.org/Publish/Author-Resources/Data-for-Authors}.

\section*{Conflict of Interest disclosure}
The authors declare there are no conflicts of interest for this manuscript.

%Please include a comprehensive conflict of interest statement that reflect all conflicts of interest for all involved authors. If there are no conflicts of interest please state, “The authors declare there are no conflicts of interest for this manuscript.”

\acknowledgments
This work is supported by the NASA Space Weather Center of Excellence program under award No. 80NSSC23M0191 and No. 80NSSC23M0192. The authors thank Ian Richardson (University of Maryland, NASA GSFC), A. Steve Johnson (Leidos, NASA JSC SRAG), Weihao Liu (University of Michigan), and Ke Hu (University of Illinois Urbana-Champaign) for their efforts in acquiring and preparing the dataset used in this study. The authors also thank the reviewers for their constructive and insightful comments, which helped improve the clarity and quality of the manuscript.

\bibliography{reference}

%Reference citation instructions and examples:
%
% Please use ONLY \cite and \citeA for reference citations.
% \cite for parenthetical references
% ...as shown in recent studies (Simpson et al., 2019)
% \citeA for in-text citations
% ...Simpson et al. (2019) have shown...
%
%
%...as shown by \citeA{jskilby}.
%...as shown by \citeA{lewin76}, \citeA{carson86}, \citeA{bartoldy02}, and \citeA{rinaldi03}.
%...has been shown \cite{jskilbye}.
%...has been shown \cite{lewin76,carson86,bartoldy02,rinaldi03}.
%... \cite <i.e.>[]{lewin76,carson86,bartoldy02,rinaldi03}.
%...has been shown by \cite <e.g.,>[and others]{lewin76}.
%
% apacite uses < > for prenotes and [ ] for postnotes
% DO NOT use other cite commands (e.g., \citeA, \cite, \citeyear, \nocite, \citealp, etc.).
%

\end{document}